\documentclass[prl,aps,showpacs,groupedaddress,twocolumn,toc=flat]{revtex4-1}
\usepackage[colorlinks=true,linkcolor=blue,urlcolor=blue,citecolor=blue]{hyperref}

\usepackage[utf8x]{inputenc}
\usepackage{color}
\usepackage{bbm} 

\usepackage{amsfonts,amsmath,amssymb,stmaryrd}

\usepackage{graphicx}
\usepackage{subfigure}  
\usepackage{bbm} 
\usepackage{hyperref}
\usepackage{epsfig}
\usepackage{mathrsfs}
\usepackage{verbatim}
\usepackage{centernot}
\usepackage{ulem}
\usepackage{xspace}
\usepackage[english, status=draft]{fixme}       
\fxusetheme{color}                              %

\usepackage{float}
\renewcommand{\l}{\left(}
\renewcommand{\r}{\right)}

\newcommand{\bra}[1]{\langle#1|}
\newcommand{\ket}[1]{|#1\rangle}

\renewcommand{\ij}{{\langle \vec{i}, \vec{j} \rangle}}
\renewcommand{\H}{\hat{\mathcal{H}}}

\renewcommand{\c}{\hat{c}^{\vphantom \dagger}}

\newcommand{\cd}{\hat{c}^\dagger}

\newcommand{\hc}{\text{h.c.}}

\usepackage{array}

\usepackage{cancel,ifthen}
\newcommand{\cmnt}[2][NoInPuT]{\ifthenelse{\equal{#1}{NoInPuT}}{}{{\color{red}\sout{#1}}} {\color{blue} #2}}

\usepackage{bm}	
\renewcommand{\vec}[1]{\bm{#1}}

\begin{document}
\normalem	

\title{Dynamical formation of a magnetic polaron in a two-dimensional quantum antiferromagnet}

\author{Annabelle Bohrdt}
\author{Fabian Grusdt}
\author{Michael Knap}
\address{Department of Physics and Institute for Advanced Study, Technical University of Munich, 85748 Garching, Germany}
\address{Munich Center for Quantum Science and Technology (MCQST), Schellingstr. 4, D-80799 M{\"u}nchen, Germany}


\date{\today}

\begin{abstract}
We numerically study the real-time dynamics of a single hole created in the $t-J$ model on a square lattice. Initially, the hole spreads ballistically with a velocity proportional to the hopping matrix element. At intermediate to long times, the dimensionality as well as the spin background determine the hole dynamics. A hole created in the ground state of a two dimensional quantum antiferromagnet propagates again ballistically at long times but with a velocity proportional to the spin exchange coupling, showing the formation of a magnetic polaron.
We provide an intuitive explanation of this dynamics in terms of a parton construction, which leads to a good quantitative agreement with the numerical simulations. In the limit of infinite temperature and no spin exchange couplings, the dynamics can be approximated by a quantum random walk on the Bethe lattice. Adding Ising interactions corresponds to an effective disordered potential, which can dramatically slow down the hole propagation, consistent with subdiffusive dynamics. 
\end{abstract}

\maketitle

\emph{Introduction.--}
Understanding the properties of a single mobile hole doped into an antiferromagnet allows one to reveal the interplay of spin and charge degrees of freedom. This constitutes a crucial step in the theoretical description of the Fermi-Hubbard model and by extension, strongly correlated cuprate compounds~\cite{Lee2006,Keimer2015}. 
\\
In the limit of strong interactions, the Fermi-Hubbard model can be approximated by the $t-J$ model. In two dimensions and for a single hole, the latter is described by the Hamiltonian~\cite{Auerbach1998},
\begin{equation}
\H_{t-J} =  -t \sum_{\ij, \sigma}\mathcal{P} \left( \cd_{\vec{i},\sigma} \c_{\vec{j},\sigma} + h.c. \right) 
\mathcal{P}+ J \sum_\ij \hat{\mathbf{S}}_{\vec{i}} \cdot \hat{\mathbf{S}}_{\vec{j}},
\label{eq:tjmodel}
\end{equation}
where $\mathcal{P}$ projects on states with less than two fermions per site. The first term describes tunneling of holes with amplitude $t$ and the second term denotes spin-exchange interactions with coupling constant $J=4 t^2/U$, where $U$ is the bare interaction. In the absence of doping, the ground state of the two dimensional system is effectively a Heisenberg antiferromagnet with long-range spin correlations. 
\\
Here we study the dynamics of the system after the creation of one hole at the origin, which can move through the antiferromagnet according to the tunneling matrix element $t$. The motion of the hole distorts the surrounding spin order and leads to an interplay of spin and charge dynamics. 
Even though this problem is most intuitively described in real space, a predominant part of research has focused on frequency and momentum space properties, which are most directly accessible in solid-state experiments. An important example is the spectral function~\cite{Dagotto1990,Liu1992,Leung1995,Brunner2000,Mishchenko2001}, which can be expressed as an overlap of the initial state with the time-evolved state with a single hole. Noteable exceptions are the works in Refs.~\cite{Zhang1991,Mierzejewski2011,Kogoj2014,Lenarcic2014,Golez2014,Eckstein2014,Carlstrom2016,KanaszNagy2017}.
\begin{figure}[ht!]
\centering
\epsfig{file=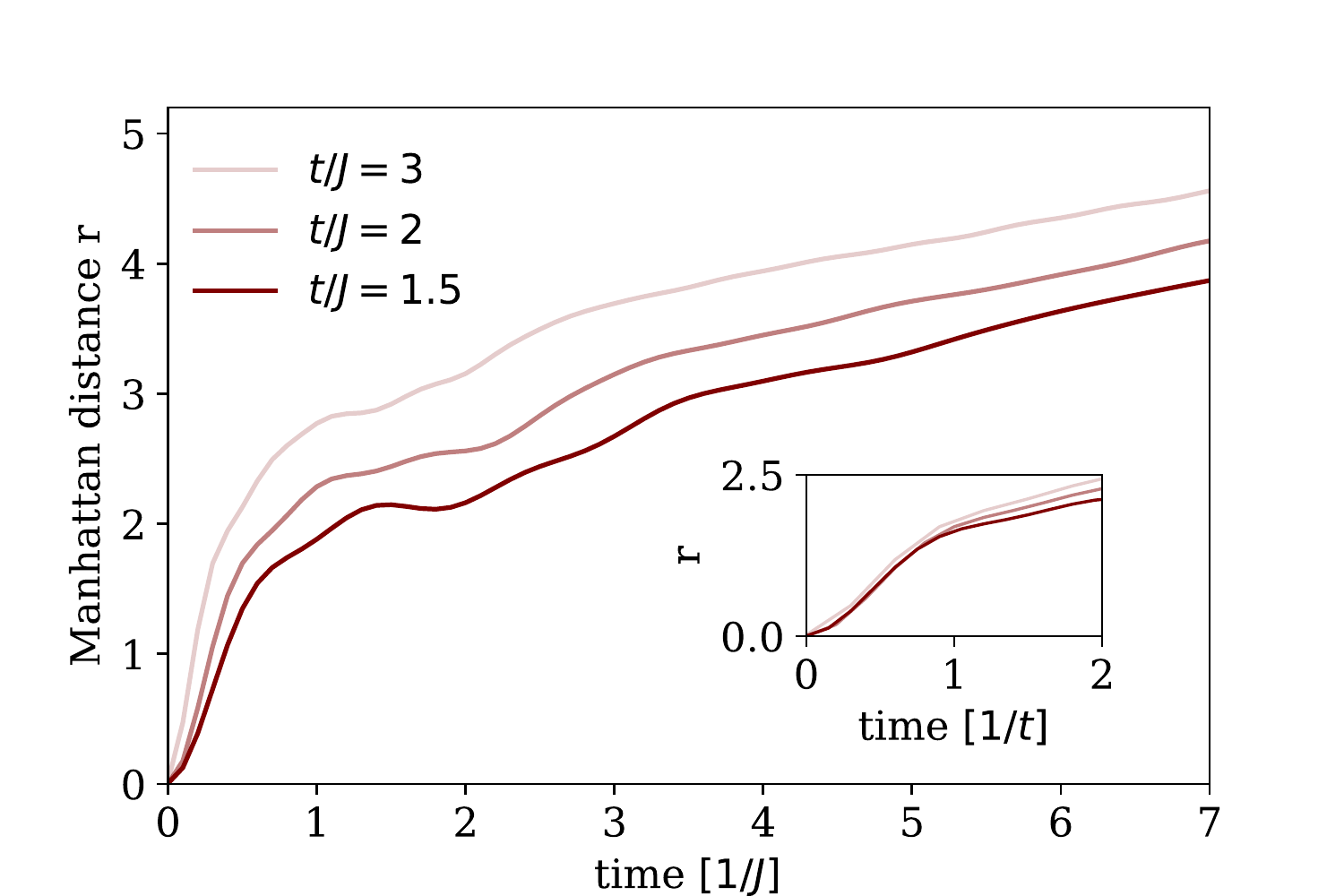, width=0.45\textwidth}
\caption{\textbf{Dynamics of a hole} in the two-dimensional $t-J$ model on a square lattice for $t/J = 1.5,2, 3$ with cylinder length $L_x=18$ and circumference $L_y = 4$. The initial spreading with velocity proportional to the hole hopping $t$ coincides for all values of the hopping for times in units of $1/t$ (inset). For longer times, the Manhattan distance increases with a slower rate determined by the spin exchange $J$ and is independent of the hopping $t$.} 
\label{fig_tDep}
\end{figure}
Creating a hole in an antiferromagnetic background realizes a local high energy excitation that is best probed with local resolution. The tremendous progress in quantum gas microscopy in recent years~\cite{Greif2013,Hart2015,Parsons2015,Cheuk2016,Mazurenko2017,Hilker2017,Brown2018,Nichols2018,Salomon2018,Chiu2018} has sparked an increasing interest in real space properties and the analysis of single snapshots obtained in projective measurements.
In particular, related experiments in one dimension have recently demonstrated spin-charge separation in space and time~\cite{Vijayan2019}.
\\
Here we use a matrix-product-operator based time evolution~\cite{Kjaell2012,Zaletel2015} on a cylinder with four legs~\cite{Gohlke2017} to efficiently calculate the dynamics of a hole introduced into a spin system according to the $t-J$ Hamiltonian (\ref{eq:tjmodel}). We find that the hole spreads differently at short and long times, Fig.~\ref{fig_tDep}.
\onecolumngrid
\vskip .6cm
\begin{figure*}
\centering
\epsfig{file=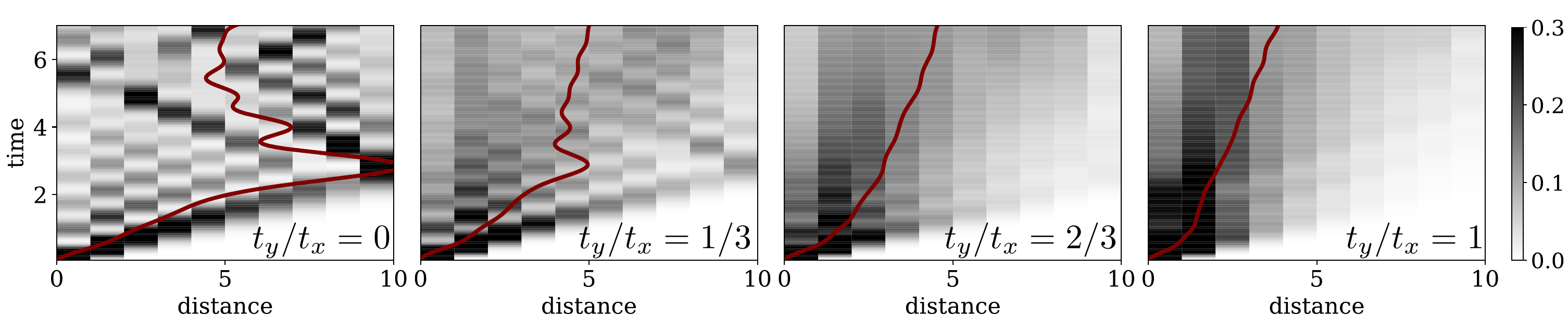, width=0.99\textwidth}
\caption{\textbf{Dimensional crossover.} Distance of the hole to the origin along the cylinder for $t_y/t_x = 0,1/3, 2/3,1$ for a cylinder of length $L_{cyl}=19$ sites. In the one-dimensional case, $t_y/t_x = 0$, the hole dynamics are coherent and the hole spreads ballistically until it reaches the edge of the cylinder. Remnants of coherent behavior are still visible for $t_y/t_x = 1/3$ for the system sizes considered here and the same boundary effects as in one dimension appear. For $t_y/t_x=2/3$ the velocity of the hole propagation is significantly reduced after the initial fast dynamics. Red lines show the mean value of the distance to the origin along the cylinder.} 
\label{figCrossover}
\end{figure*}
\twocolumngrid
At short times, the hole propagates ballistically with a velocity proportional to $t$, see inset of Fig.~\ref{fig_tDep}.
 However, at a time $~1/J$, the propagation slows down. 
We explain our numerical findings by a parton construction, where the excitation created in the system is decomposed into a spinon, carrying the spin quantum number, and a chargon, carrying the charge~\cite{Beran1996,Manousakis2007,Grusdt2018tJz,Grusdt2019}. The propagation of the hole at intermediate and long times is therefore dominated by the slow spinon dynamics. The ballistic spreading with a velocity proportional to the spin exchange $J$ at long times is due to the finite quasiparticle weight of the magnetic polaron.
\\

\emph{Dimensional crossover from 1D to 2D.--}
The constituents used to describe excitations in the $t-J$ model carry either spin (spinons) or charge (chargon). In one spatial dimension spin-charge separation occurs; spinon and chargon are deconfined and propagate with different velocities~\cite{Vijayan2019, giamarchi}.
\\
In one dimension, the independent dynamics of the quasiparticles can be understood in squeezed space, where sites occupied by a hole are removed from the lattice, such that the system is described by the combined information of hole positions and spin configuration~\cite{Kruis2004a,Hilker2017}. Most importantly, the motion of a hole cannot change the relative positions of the spins in one dimension.
\\
In a cold atom experiment realizing the Fermi-Hubbard model, the crossover from one to two dimensions can be studied by tuning the lattice depth in one spatial direction~\cite{Salomon2018}. In particular, the interaction $U$ stays constant while the ratio $t_y/t_x$ is tuned, and upon mapping the Fermi-Hubbard model to the $t-J$ model, the spin exchange couplings become $J_y/J_x = t_y^2/t_x^2$. 
\\
In Fig.~\ref{figCrossover} the dynamics of the hole is studied for different values of $t_y/t_x$. For $t_y/t_x=0$, we observe coherent hole dynamics as expected from spin-charge separation.
For finite transverse couplings $t_y/t_x >0$, the one-dimensional picture of spin-charge separation breaks down. While at short times, the spreading of the hole along the finite size cylinder still resembles the coherent dynamics of $t_y/t_x=0$, at longer times, the mean distance to the origin increases significantly slower in the case of finite $t_y/t_x$ (red lines). 
\\
The qualitative change in the hole dynamics can be understood from a parton construction of spinons and chargons~\cite{Grusdt2019}. Once the hopping in the second dimension is turned on, the hole motion distorts the spin background and frustrates the antiferromagnetic correlations of the system. This energy cost increases as the chargon hops away from the spinon and the chargon motion is therefore associated with a potential energy cost for the spin system. The competition between kinetic energy of the hole and potential energy of the spin system leads to an emergent length scale, which can be observed in Fig.~\ref{figCrossover} as the distance at which the velocity of the propagation changes. 


\emph{Chargon and spinon dynamics in 2D.--}
\begin{figure}
\centering
\epsfig{file=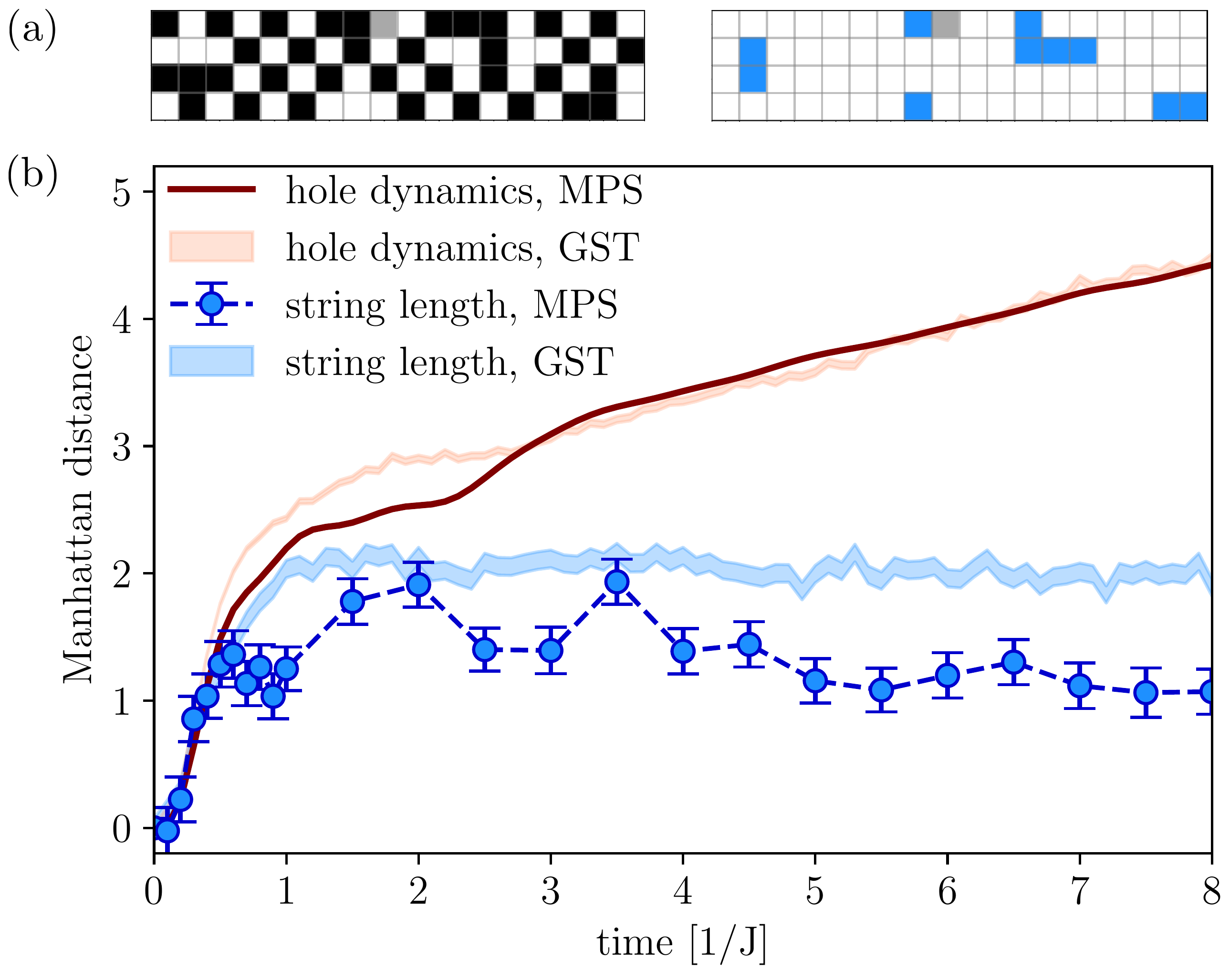, width=0.46\textwidth}
\caption{\textbf{A parton interpretation of the hole dynamics} for $t/J = 2$. (a) In the geometric string theory, spinon and chargon are bound together by a string of displaced spins. We extract string patterns by comparing to the perfect N\'eel state. In each snapshot, starting from the hole, the longest deviation from the N\'eel state without off-branches is taken as string pattern. (b) Manhattan distance of the hole to the origin and average string pattern length obtained from matrix product state (MPS) snapshots are compared to geometric string theory (GST) predictions. The latter are obtained based on snapshots from the ground state of the Heisenberg model.} 
\label{fig2D}
\end{figure}
For $t_y=t_x=t$, we observe a separation of time scales in the hole dynamics, see Fig.~\ref{fig_tDep}. The hole shows fast ballistic spreading initially until a time $\approx 1/t$, at which its velocity is drastically reduced. At longer times $\approx 1/J$, the distance of the hole to the origin continues to increase linearly but at a slower rate. Our analysis for different values of $t/J$ in Fig.~\ref{fig_tDep} demonstrates that the initial dynamics are universal and do not depend on the value of $J$. At long times we observe slow spreading with a velocity essentially independent of $t$. 
\\
Within the geometric string theory~\cite{Grusdt2018tJz,Grusdt2019}, this behavior can be understood as follows. The spinon and the chargon are bound to each other by a string of displaced spins, where the length of this string is determined by the string tension~\cite{Grusdt2019}. The string tension for a straight string is given by
\begin{equation}
\frac{\text{d}E_{\text{pot}}}{\text{d}l} = 2J \cdot \left( C_{\mathbf{e}_x+\mathbf{e}_y} - C_{\mathbf{e}_x}\right),
\end{equation}
because spins which were diagonal next nearest neighbors before become nearest neighbors. 
Here, $C_d = \left<\psi_0 \right| \mathbf{\hat{S}}_\mathbf{d} \cdot  \mathbf{\hat{S}}_\mathbf{0} \left| \psi_0 \right>$ and $\left| \psi_0 \right>$ is the initial state before the creation of the hole, see~\cite{supp}. 
Assuming a frozen spin background, the backaction onto the spin states is neglected and the chargon moves in a frame relative to the spinon. Up to self-retracing parts, all paths of the chargon are assumed to be distinguishable since they lead to different spin configurations in almost all cases. The chargon therefore effectively moves on a Bethe lattice and the string binding the chargon to the spinon experiences an approximately linear potential. 
\\
After the initial spreading out, a further motion of the hole is restricted and the Manhattan distance to the origin reaches a plateau. For longer times the dynamics of the spinon itself, introduced through spin-exchange processes on a time scale of $1/J$, becomes important. Within the geometric string theory, this corresponds to a motion of the origin of the Bethe lattice defined by the string states~\cite{supp}. We thus observe a ballistically propagating hole, which corresponds to a heavily dressed magnetic polaron~\cite{Koepsell2018} with finite quasiparticle weight at zero temperature with a bandwidth proportional to $J$.
\\
The quantum many-body state of the system is a superposition of many different Fock space configurations. When evaluating conventional observables such as the hole density, we thus average over a variety of string lengths and configurations. However, similar to measurements in a quantum gas microscope, we can numerically sample Fock space snapshots from the time evolved matrix product state. In each snapshot, we extract the string pattern as the longest connected deviation from the perfect N\'eel state originating from the hole~\cite{Chiu2018,Grusdt2019}, see Fig.~\ref{fig2D}a). Since the ground state of the $t-J$ model itself already deviates from the N\'eel state, even at the starting point of the dynamics the average string pattern length is finite. We thus consider the difference of the average string pattern length to its value at time zero. The string pattern length grows at short times exactly as the distance of the hole to its initial point, see Fig.~\ref{fig2D}b). As the velocity of the hole changes, the average string pattern length reaches a constant value, indicating the existence of a string which prevents a further separation of spinon and chargon.
\\
Based on snapshots of the Heisenberg groundstate, we can use the predictions of the geometric string theory in combination with a simple theory for the spinon dynamics to generate a new dataset for each time step. Using a spinon dispersion extracted from Ref.~\cite{Martinez1991}, see also~\cite{supp}, we first sample the position of the spinon. We then place a hole at this site and move it through the system for a number of steps sampled from the string length distribution given by the geometric string theory, where in each step the direction is chosen randomly. From this new dataset, the distance of the hole to the center of the cylinder as well as the average length of the string pattern can be extracted as before. For both quantities, geometric string theory combined with spinon dynamics yields good agreement with the numerical simulations, see Fig.~\ref{fig2D}b). At long times, the string pattern length extracted from the matrix product state snapshots decays slightly. We attribute this to interactions of the string with the spin environment, leading to an effective decay of the string length. In the geometric string theory, we do not take such effects into account, and thus the string pattern length stays constant after times $\propto 1/J$.
\\

\begin{figure}[t!]
\centering
\epsfig{file=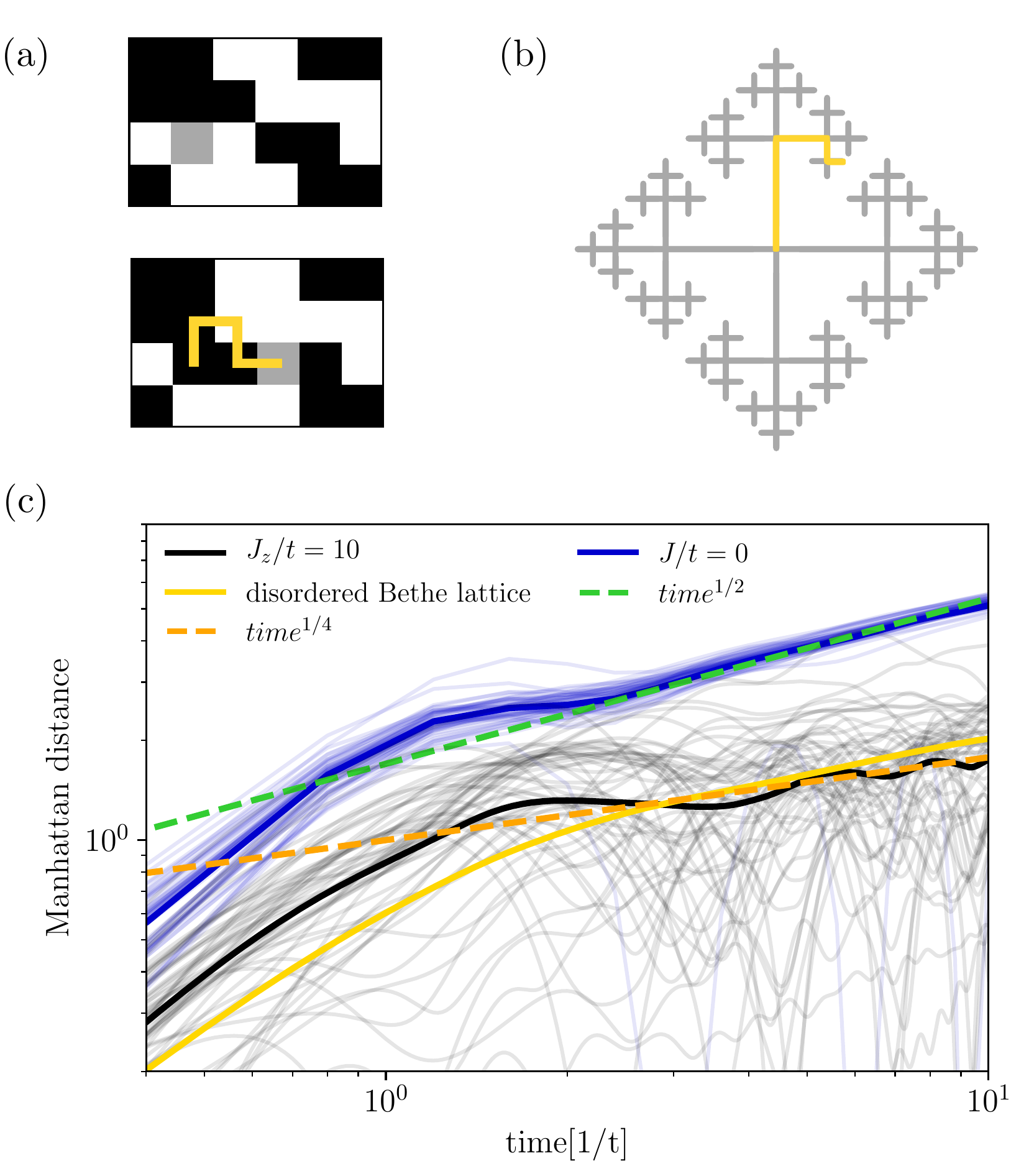, width=0.49\textwidth}
\caption{\textbf{Dynamics of a hole in an infinite temperature initial state.} (a) A hole is created in the center of a cylinder with a random configuration of spins. As the hole moves through the system, it rearranges the spins and thereby leaves a memory. (b) Its dynamics can therefore be approximated as a quantum random walk on the Bethe lattice. (c) For no spin couplings, $J=0$, the hole spreads diffusively at intermediate times. In the case of strong Ising interactions, $J_z = 10t$, the spreading is limited to comparably short distances. The faint lines are the results for individual inital product states. Fluctuations strongly increase for finite $J_z$. The yellow line corresponds to the dynamics of a single particle on the Bethe lattice with a disorder potential.}
\label{figInfiniteT}
\end{figure}

\emph{Infinite temperature.--}
In quantum gas microscopy experiments, the temperature is always finite, with the lowest achieved values to date of $T/J=0.5$.  
The results presented so far started from the ground state of the spin system, where long-range spin correlations are present. At finite temperature, the chargon dynamics should be well described by the geometric string theory as long as the short range spin-correlations and thereby the string tension are finite.
\\
The limit of infinite temperature can be studied by considering an ensemble of random product states. In this case, all spin correlations are zero and correspondingly the string tension vanishes. Nonetheless, the hole motion is associated with a memory effect in the spin system, since in general different paths taken by the hole lead to different spin configurations. In the case of $J\approx 0$, the hole dynamics can be approximated by a quantum random walk on the Bethe lattice~\cite{Brinkman1970}, assuming that all paths are distinguishable. Mapping back to the square lattice, this translates to diffusive behavior~\cite{KanaszNagy2017}. For the spin $1/2$-system considered here, different paths of the hole can lead to the same spin configuration. As a consequence, the diffusive behavior expected from the Bethe lattice calculation will be slightly modified~\cite{KanaszNagy2017}. As shown in Fig.~\ref{figInfiniteT}, the Manhattan distance of the hole to its initial point is consistent with diffusive behavior.
\\
Adding Ising interactions effectively creates a disorder potential on the Bethe lattice: in each Fock space configuration, the motion of the hole by one site from $i$ to $j$ changes the energy of the spin system by $\Delta \epsilon_{\left<ij\right>} = 0.25 J^z \left(\Delta N_{\sigma \bar{\sigma}} - \Delta N_{\sigma \sigma}  \right)$, where $\Delta N_{\sigma (\bar{\sigma})\sigma}$ is the change in the number of (anti-)aligned spins on neighboring sites. The energy difference $\Delta \epsilon_{\left<ij\right>}$ is therefore a random number between $\pm 0.5 (z-1) J^z$ with $z$ the coordination number of the lattice. The hole motion can then be approximated by a quantum random walk on the Bethe lattice with a disorder potential $W_{l} = \sum_{\left<ij\right> \in \Sigma} \Delta \epsilon_{\left<ij\right>}$, where the sum runs over all bonds $\left<ij\right>$ along the string $\Sigma$. 
For the spin configuration depicted in Fig.~\ref{figInfiniteT}a), the potential of the different hole position along the considered path is $W_l/J_z = 0, -1.5, 0, 0.5,0.5$. Note that for sites further apart on the Bethe lattice, the range of possible energy differences scales with the distance between the sites.  
In the case of strong Ising interactions, $J_z =10t$, the spreading of the hole is significantly reduced and consistent with subdiffusive spreading, see Fig.~\ref{figInfiniteT}b). 
\\

\emph{Summary and Outlook.--}
In this work, we have studied the dynamics of a single hole created in a spin system. 
Our findings at zero temperature can be understood in the framework of a parton construction, where the original excitation is described by a chargon and a spinon connected by a string of displaced spins. For infinite temperature and no spin couplings, we confirm indications for diffusive behavior~\cite{Carlstrom2016,KanaszNagy2017} for longer times. This can be explained by the dynamics of a free particle on the Bethe lattice. Introducing Ising interactions adds a disorder potential, leading to hole dynamics consistent with subdiffusion in the limit of $J_z \gg t$.
\\
In the future, it would be interesting to study the effect of finite temperature $0<T<\infty$ on the observed behavior. For experimentally realistic temperatures of $T/J \approx 0.5$, the spin correlation functions, especially on short distances, are still sizeable and thus the string tension is finite. The inverse temperature sets a timescale on which we expect a crossover from ballistic to diffusive behavior.
\\
The case of infinite temperature and large Ising interactions is approximated by a particle moving on the Bethe lattice with a disordered potential, where the strength increases with the depth in the lattice. This scenario could be studied in more detail theoretically and possibly experimentally, for example with Rydberg interactions~\cite{Grusdt2018tJz}.
\\
Starting from the ground state or low temperatures, the same analysis for the time evolution of the hole distance as well as the string pattern length could be applied in the case of adiabatically releasing the hole. 
\\
Another exciting direction for future research is the study of finite doping, especially the case of two holes. 
In the spirit of the analysis of single snapshots, it would be intriguing to see if string patterns connecting the two holes can be found or if they form two separate bound objects with a spinon each. 
\\

\emph{Acknowledgements.--}
We would like to acknowledge fruitful discussions with Immanuel Bloch, Christie Chiu, Eugene Demler, Markus Greiner, Geoffrey Ji and Guillaume Salomon. We additionally want to thank Frank Pollmann and Ruben Verresen for providing important parts of the numerical code. 
We acknowledge support from the Technical University of Munich - Institute for Advanced Study, funded by the German Excellence Initiative and the European Union FP7 under grant agreement 291763, the Deutsche Forschungsgemeinschaft (DFG, German Research Foundation) under Germany's Excellence Strategy--EXC-2111--390814868, DFG grant No. KN1254/1-1, DFG TRR80 (Project F8), and the Studienstiftung des deutschen Volkes.

\bibliography{HoleDyn.bib}
\bibliographystyle{unsrt}

\clearpage
\newpage
\section*{Supplementary information}

\subsection{Models}

In the main text, besides the $t-J$ model itself, generalizations thereof are considered. 
For the dimensional crossover, the couplings along and around the cylinder are assumed to be independently tunable,
\begin{multline}
\H_{t-J} =  - \sum_{\vec{i}, \sigma} \mathcal{P}  \left(  t_x \cd_{\vec{i},\sigma} \c_{\vec{i}+\vec{e_x},\sigma} +   t_y  \cd_{\vec{i},\sigma} \c_{\vec{i}+\vec{e_y},\sigma}+\hc \right) \mathcal{P} \\
+
\sum_{\vec{i}} J_x \hat{\mathbf{S}}_{i} \cdot \hat{\mathbf{S}}_{\vec{i}+\vec{e_x}} + J_y \hat{\mathbf{S}}_{i} \cdot \hat{\mathbf{S}}_{\vec{i}+\vec{e_y}} .
\label{eq:crossover}
\end{multline}

In the case of an infinite temperature initial state, the coupling constant of the spin exchange part of the Hamiltonian is tuned independently for the Ising part of the interactions, assuming a $t-\rm XXZ$ Hamiltonian:  
\begin{multline}
\H_{t-{\rm XXZ}} =  -t \sum_{\ij, \sigma} \mathcal{P} \left( \cd_{i,\sigma} \c_{j,\sigma} + \hc \right) \mathcal{P} + J_z \sum_{\ij} \hat{S}_{i}^z \hat{S}_j^z \\
+ \frac{J_\perp}{2} \sum_{\ij}\left( \hat{S}_{i}^+\hat{S}_j^- + \hc \right)  ,
\label{eq:tjzmodel}
\end{multline}

\subsection{Numerics}

\subsubsection{Snapshots}
We obtain Fock space snapshots of the time evolved state of the system by Metropolis Monte Carlo sampling from the matrix product state. Initially, a random configuration is chosen as state $\ket{\psi_A}$, where the number of spin ups, spin downs and holes is fixed and cannot change during the sampling procedure. Next, an update is proposed, in which the state $\ket{\psi_B}$ is obtained from $\ket{\psi_A}$ by exchanging the occupation of two sites, which can both be located anywhere in the system. The probability $p_{A(B)}$ for state $\ket{\psi_{A(B)}}$ is given by its overlap with the matrix product state under consideration. The new state $\ket{\psi_B}$ is thus accepted as new state with a probability given by the ratio $p_B/p_A$. The process is repeated until the probability does not change significantly any more. Then, the current state is saved as new snapshots every 1000 steps.

\subsubsection{Convergence}
\begin{figure}
\centering
\epsfig{file=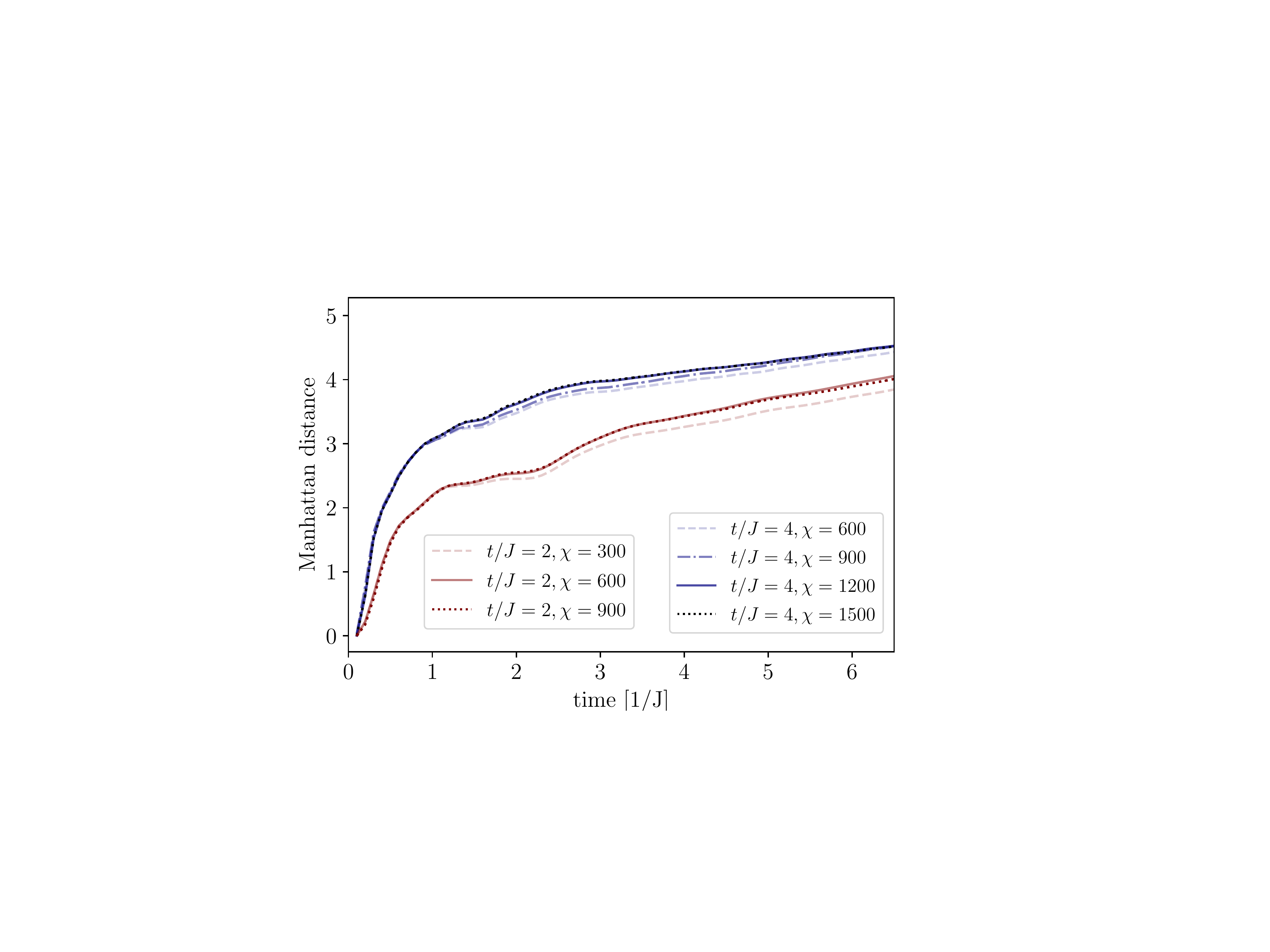, width=0.49\textwidth}
\caption{\textbf{Convergence with bond dimension} of the Manhattan distance of the hole for $t/J=2$ and $t/J=4$.  } 
\label{figBondDim}
\end{figure}

The results shown in the main text were obtained using the TenPy package~\cite{Kjaell2012,Zaletel2015,Hauschild2018} and have been carefully checked for convergence in various parameters such as the bond dimension. As shown in Fig.~\ref{figBondDim}, the Manhattan distance of the hole to the origin does not change with increasing bond dimension for $\chi > 600$ for $t/J=2$ and $\chi > 1000$ for $t/J=4$. 

\subsubsection{System size}

\begin{figure}
\centering
\epsfig{file=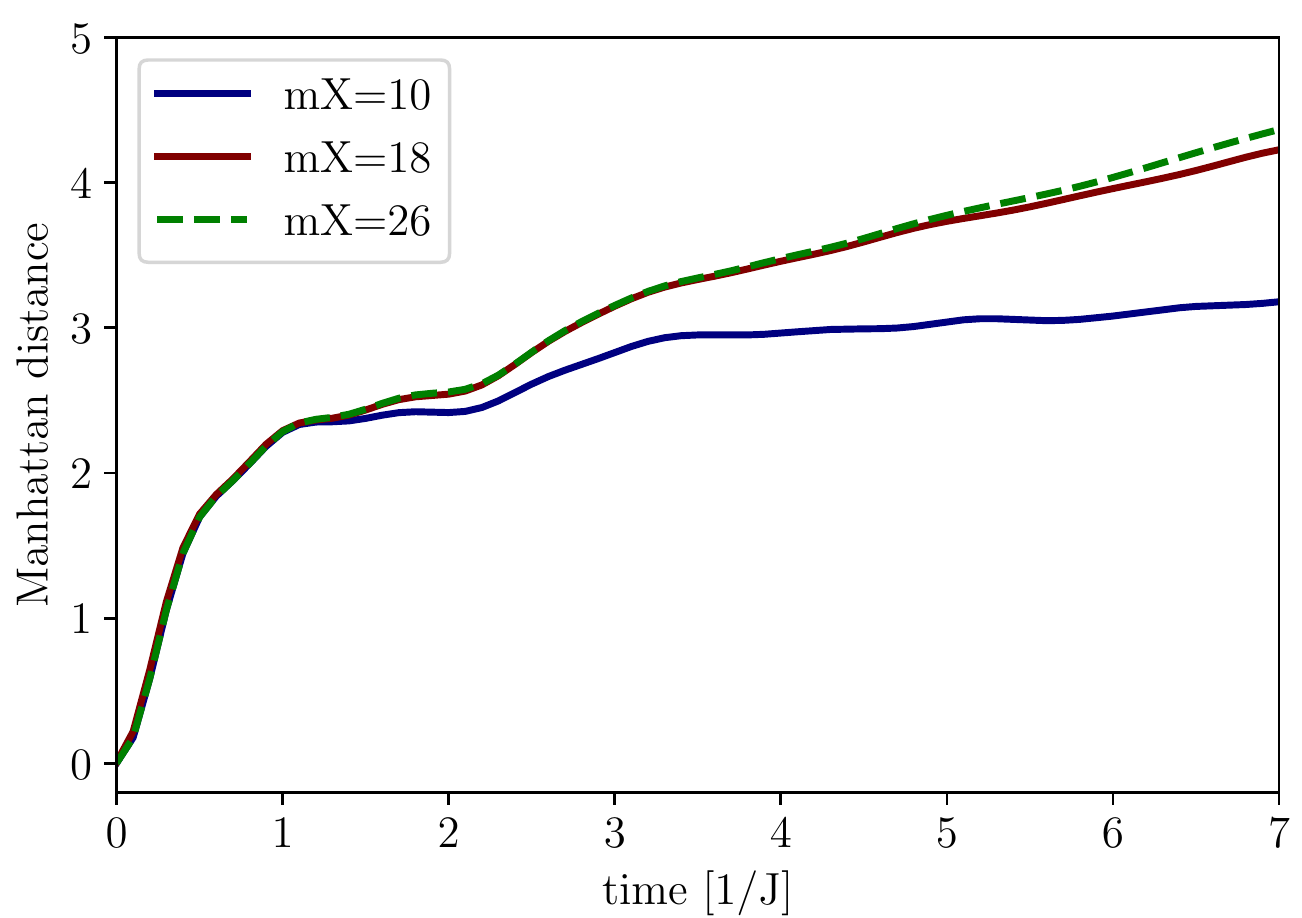, width=0.49\textwidth}
\caption{\textbf{System size dependence} of the Manhattan distance of the hole for $t/J=2$. For a cylinder of length 10 sites, finite size effects become visible at an earlier time. The dynamics of the observed quantities is very similar for cylinders of length 18 and 26 sites up to longer times.} 
\label{figShort}
\end{figure}

The system sizes considered here are finite and at long times, saturation of the Manhattan distance of the hole occurs. The corresponding time scales depend on the size of the system. As shown in Fig.~\ref{figShort}, the distance of the hole to the origin already saturates before the ballistic spread with a second velocity becomes visible. However, the dynamics for a cylinder of length 18, which is considered in the main text, coincides with the dynamics for a cylinder of length 26 sites up to times $\approx 5/J$. At longer times, the Manhattan distance increases a bit faster for the longer system, indicating that effects from the finite cylinder length become visible for the 18 site long cylinder.

\subsection{Theory}

In this section we describe the parton theory of hole dynamics to which we compare our numerical results in the main text. The theoretical formalism used here has been introduced previously in~\cite{Grusdt2018tJz,Grusdt2019,Chiu2018,Koepsell2018}, where more details can be found. 

Our basic assumption is that the Hilbert space can be approximated as a tensor product,
\begin{equation}
\mathscr{H} = \mathscr{H}_{\rm sp} \otimes \mathscr{H}_\Sigma,
\end{equation}
where $\mathscr{H}_{\rm sp}$ is the space spanned by all one-spinon states and $\mathscr{H}_\Sigma$ represents the Hilbert space of geometric strings. Because every string can be represented by a site on the Bethe lattice with coordination number $z=4$, $\mathscr{H}_\Sigma$ is equivalent to a single particle hopping between the sites of this Bethe lattice. A state $\ket{\vec{j}^{\rm s},\sigma}\ket{\Sigma}$ in the approximate Hilbertspace is associated with the following state in the $t-J$ model (up to normalization):
\begin{equation}
\ket{\vec{j}^{\rm s},\sigma}\ket{\Sigma} \sim \ket{\Psi(\Sigma,\vec{j}^{\rm s},\sigma)} = \hat{G}_\Sigma \hat{c}_{\vec{j}^{\rm s},\sigma} \ket{\Psi_0}.
\end{equation}
Here $\ket{\Psi_0}$ denotes the undoped ground state of the spin-exchange part of the $t-J$ Hamiltonian, $\hat{c}_{\vec{j}^{\rm s},\sigma}$ annihilates a fermion with spin $\sigma$ on site $\vec{j}^{\rm s}$ and $\hat{G}_\Sigma$ creates a geometric string by moving the hole and displacing spins along $\Sigma$ starting from $\vec{j}^{\rm s}$. 

The effective Hamiltonian in the Hilbertspace $\mathscr{H}$ can be constructed by calculating the matrix elements $\bra{\Psi(\Sigma_2,\vec{j}^{\rm s}_2,\sigma_2)} \H_{t-J} \ket{\Psi(\Sigma_1,\vec{j}^{\rm s}_1,\sigma_1)}$ of the $t-J$ Hamiltonian $\H_{t-J}$ in the physical Hilbertspace. This leads to direct hopping with amplitude $t$ between neighboring sites of the Bethe lattice, corresponding to different string configurations, $\langle \Sigma_2,\Sigma_1 \rangle$, 
\begin{equation}
\H_{\Sigma} = - t \sum_{\langle \Sigma_2,\Sigma_1 \rangle } \biggl( \ket{\Sigma_2} \bra{\Sigma_1} + \hc \biggr),
\end{equation}
a tight-binding Hamiltonian of spinons
\begin{multline}
\H_{\rm sp} = \sum_{\sigma} \sum_{\vec{i}, \vec{j}}  J_{\vec{j}, \vec{i}} \l  \hat{s}^\dagger_{\vec{j},\sigma} \hat{s}_{\vec{i},\sigma} + \hc  \r \\
= \sum_{\sigma} \sum_{\vec{k}}  \omega_{\rm sp}(\vec{k})  \hat{s}^\dagger_{\vec{k},\sigma} \hat{s}_{\vec{k},\sigma} ,
\label{eqHspinon}
\end{multline}  
 and a potential energy term 
\begin{equation}
\H_{\rm pot} = \sum_\Sigma \ket{\Sigma} \bra{\Sigma} ~ V_J(\Sigma), 
\end{equation}
where $V_J(\Sigma) = \bra{\Psi(\Sigma,\vec{j}^{\rm s},\sigma)} \H_{J} \ket{\Psi(\Sigma,\vec{j}^{\rm s},\sigma)}$. We neglect other matrix elements, see Ref.~\cite{Grusdt2018tJz} for a discussion.

The effective Hamiltonian is defined in the approximate Hilbertspace,
\begin{equation}
\H_{\rm eff} = \H_{\Sigma} + \H_{\rm pot} + \H_{\rm sp},
\end{equation}
and does not couple the spin and charge (string) sectors. This allows for a simple solution of the quench dynamics described in the main text. Starting from the initial state $\ket{\Phi(0)} = \sum_{\sigma} \ket{\vec{j}^{\rm s}=0,\sigma} \ket{\Sigma=0}$ we obtain a solution in the form of a product state:
\begin{equation}
\ket{\Phi(t)} = e^{- i \H_{\rm sp} t}  \sum_{\sigma} \ket{\vec{j}^{\rm s}=0,\sigma} e^{- i(\H_\Sigma + \H_{\rm pot}) t} \ket{\Sigma=0}.
\end{equation}

We can further simplify the effective Hamiltonian by making the linear string approximation. There we assume that $V_J(\Sigma) \propto \ell_\Sigma$ is proportional to the length $\ell_\Sigma$ of the string and does not depend on its orientation. This condition is not exactly satisfied, but it provides a good approximate description \cite{Grusdt2018tJz}. An expression for $V_J(\ell_\Sigma)$ can be obtained by considering only straight strings:
\begin{equation}
V_J(\ell_\Sigma) = \frac{dE}{d\ell} \times \ell_\Sigma + g_0 \delta_{\ell_\Sigma,0} + \mu_{\rm h},
\end{equation}
where $dE / d\ell = 2 J (C_2 - C_1)$, $g_0 = - J (C_3 - C_1)$ and $\mu_{\rm h} = J (1 + C_3 - 5 C_1)$ with $C_1 = \bra{\Psi_0} \hat{S}_{\vec{e}_x} \cdot \hat{S}_{\vec{0}} \ket{\Psi_0}$, $C_2 = \bra{\Psi_0} \hat{S}_{\vec{e}_x + \vec{e}_y} \cdot \hat{S}_{\vec{0}} \ket{\Psi_0}$ and $C_3 = \bra{\Psi_0} \hat{S}_{2 \vec{e}_x} \cdot \hat{S}_{\vec{0}} \ket{\Psi_0}$ denoting nearest, next-nearest and next-next-nearest neighbor correlations in the undoped spin system \cite{Grusdt2019}.

\begin{figure}
\centering
\epsfig{file=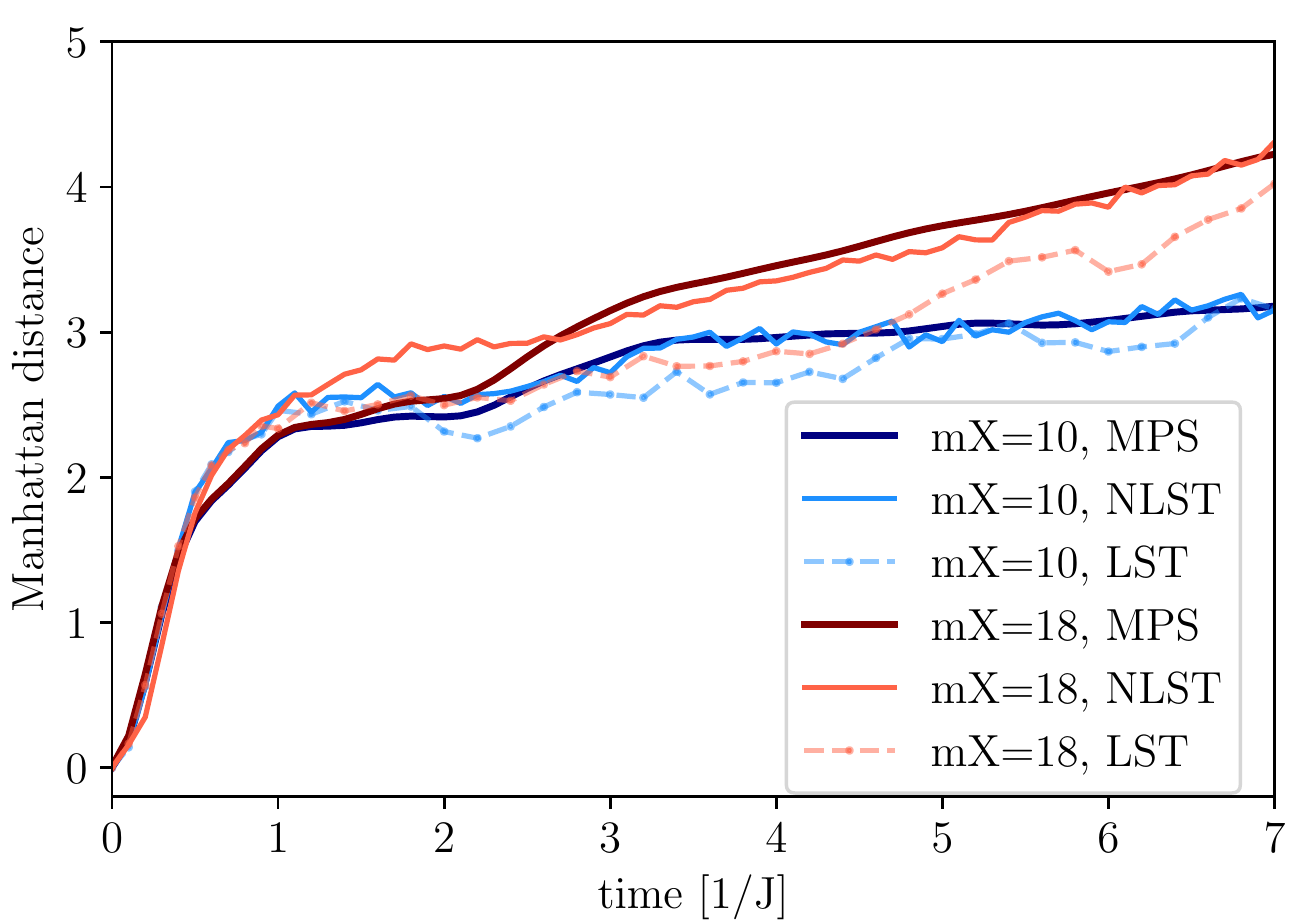, width=0.49\textwidth}
\caption{\textbf{Linear versus non-linear string theory.} Comparison of the two theoretical approximations to the numerical results for the Manhattan distance of the hole. Non-linear string theory results are limited to a string length of 10 bonds. } 
\label{figNLST}
\end{figure}

For the theory comparisons shown in the main text, the non-linear string theory was used. 
In Fig.~\ref{figNLST}, we compare the numerical simulation to linear as well as non-linear string theory predictions for cylinders of length $L_{cyl}=10$ and $L_{cyl}=18$. In both cases, the non-linear string theory agrees significantly better with the numerical results. For the string length distribution used in this case, we are restricted to a maximum string length of 10 bonds due to the size of the corresponding Bethe lattices. In the linear string theory, we can make use fo the rotational symmetries of the Bethe lattice. Thus, the string dynamics can be solved for maximum string lengths as large as desired. We solve the single spinon problem on a cylinder with the same circumference as used in the exact numerics in both cases.

We determine the spinon dispersion $\omega_{\rm sp}(\vec{k})$ in Eq.~\eqref{eqHspinon}, and thereby the spinon tunneling matrix elements $J_{\vec{j},\vec{i}}$ in the effective Hamiltonian, by fitting it to the magnetic polaron dispersion:
\begin{multline}
\omega_{\rm sp}(\vec{k}) = A \left[  \cos( 2 k_x ) + \cos( 2 k_y ) \right]  \\
+ B \left[  \cos( k_x + k_y ) + \cos(  k_x - k_y ) \right].
\end{multline}
The fit parameters $A$ and $B$ depend weakly on $t/J$, and we take them from Ref.~\cite{Martinez1991}. For $t/J = 2$ this yields: $A=0.25 J$ and $B=0.36 J$.


\end{document}